\documentclass[useAMS,usenatbib]{mn2e}
\usepackage{epsfig}
\usepackage{graphicx}

\title[Gamma-ray burst rate]{Gamma-ray burst rate: high-redshift excess and its possible origins}
\author[F. J. Virgili, B. Zhang, K. Nagamine and J.-H. Choi]{Francisco J. Virgili$^{1}$\thanks{E-mail:
virgilif@physics.unlv.edu (FJV); zhang@physics.unlv.edu (BZ); kn@physics.unlv.edu (KN)}, Bing Zhang$^{1}$ Kentaro Nagamine$^{1}$ and Jun-Hwan Choi$^{2}$\\ 
$^{1}$Department of Physics and Astronomy, University of Nevada Las Vegas, 4505 Maryland Parkway, Las Vegas, NV 89154, USA\\
$^{2}$Department of Physics and Astronomy, University of Kentucky, 600 Rose Street, Lexington, KY, 40506}
\begin{document}

\date{Accepted xxxx. Received xxxxx; in original form xxxxx}

\pagerange{\pageref{firstpage}--\pageref{lastpage}} \pubyear{2002}

\maketitle

\label{firstpage}

\begin{abstract}
Prompted by various analyses of long (Type II) GRB rates and their relationship to the cosmic star-formation history, metallicity and luminosity function evolution, we systematically analyse these effects with a Monte Carlo code.  We test various cosmic star-formation history models including analytical and empirical models as well as those derived from cosmological simulations.  We also explore expressions for metallicity enhancement of the GRB rate with redshift, as presented in the literature, and discuss improvements to these analytic expressions from the point of view of galactic evolution.  These are also compared to cosmological simulations on metal enrichment.  Additionally we explore possible evolutionary effects of the GRB rate and luminosity function with redshift.  The simulated results are tested with the observed \textit{Swift} sample including the $L$, $z$, and peak flux ($\log N-\log P$) distributions.  The observational data imply that an increase in the GRB rate is necessary to account for the observations at high redshift, although the form of this enhancement is unclear.  A rate increase due to lower metallicity at higher redshift may not be the singular cause and is subject to a variety of uncertainties.  Alternatively, evolution of the GRB luminosity function break with redshift shows promise as a possible alternative.
\end{abstract}


\section{Introduction}

Since firmly establishing the cosmological nature of Type II (long-soft) gamma-ray bursts (GRBs) (\cite{metzger97, vanparadijs97}), there have been many predictions as to how early in cosmic history GRBs are created.  Redshifts for GRBs have been detected more effectively since the 2004 launch of the \textit{Swift} satellite (\cite{gehrels04}) which has the advantage of providing prompt localizations.  This combined with the dedicated work of ground-based astronomers has shown progress in pushing toward the theoretical detection limit of about a $z \sim 20$ (\cite{abel02,bromm02}).  Record-breaking bursts, such as 050904 ($z=$6.3, \cite{cusumano06,haislip06,kawai06,frail06}), 080913 ($z=$6.7, \cite{greiner09}) and 090423 (z=8.2, \cite{salvaterra09, tanvir09}) demonstrate just how far these objects can be detected and warrant a discussion on how bursts that occur at such drastically different times in the evolution of the universe may or may not differ. 

It is believed that Type II\footnote{See \cite{zhang09} for a full discussion on the classification of GRBs and a full description of the distinction between Type I and Type II bursts.} GRBs are a product of the core-collapse of massive stars, stemming from the evidence of an association of these GRBs with core-collapse supernovae (\cite{stanek03,hjorth03}).  These observations lead naturally to the expectation that the rate of these objects follow the cosmic star-formation history (SFH) (\cite{wijers98, totani99, lambandreichart00, blainandnat00, pandm01}).  Various studies have shown that the rate of GRBs does not strictly follow the SFH but is actually enhanced by some mechanism at high-$z$ (\cite{daigne06,le07, guetta07, li08, kistler08, kistler09, salvaterra09,salvaterra09b,salvaterra07, qin10, wanderman10,campisi10}), be it metallicity effects (\cite{langerandnorm06, li08}), selection effects or an increase in luminosity.  

In this analysis we combine and expand various elements from these works to further analyse possible GRB rate enhancements with redshift and the underlying causes and forms of these evolutions using the available observational data together with a Monte Carlo code.  We look into the underlying form of the cosmological SFH, including models derived from cosmological smooth-particle hydrodynamic (SPH) simulations (\cite{choiandnag10}), metallicity effects, rate evolution with redshift and evolution of the break luminosity of the GRB luminosity function.  

In Section 2 we present and explain the details of the various simulations that were conducted, broken down by the form of the SFH or high-$z$ enhancement (e.g. metallicity or evolution effect).  Section 3 explains the method of testing for consistency.   Section 4 details the results for the simulations in the same framework as Section 2, dedicating a section to each form of enhancement.  We conclude with a summary and discussion in Section 5. 

\section{Theory and simulations}

One of the major goals of this and previous analyses is to constrain the \textit{intrinsic} distribution of GRBs by utilizing the available \textit{observed} data.  We develop a Monte Carlo code that randomly creates a set of GRBs, defined by a luminosity and redshift pair from assumed luminosity and redshift distributions, and cycles them through a series of filters that act as a `detection'.  The setup is similar to that of \citet{virgili09} but with various additions and improvements tailored to this specific problem.  The set of generated bursts is then compared to the current observations of the luminosity, redshift, and peak flux (i.e. $\log N - \log P$) distributions.  The observed GRB rate follows the form
\begin{equation}
\frac{dN}{dtdzdL}=\frac{R_{GRB}(z)}{1+z}\frac{dV(z)}{dz}\Phi(L),
\label{dn}
\end{equation}
where the $(1+z)$ factor accounts for the cosmological time
dilation, $R_{GRB}(z)$ is the GRB volume event rate (in units of 
${\rm Gpc^{-3}~yr^{-1}}$) as a function of $z$, $\Phi(L)$ is
the luminosity function, and $dV(z)/dz$ the comoving volume element
given by
\begin{equation}\label{volume}
\frac{dV(z)}{dz}=\frac{c}{H_{\rm 0}}\frac{4\pi D_{L}^2}{(1+z)^2
[\Omega_M(1+z)^3+\Omega_\Lambda]^{1/2}},
\end{equation}
for a flat $\Lambda$ cold dark matter ($\Lambda$CDM) universe.
We assume $H_{\rm0}=\rm71~km~s^{-1}~Mpc^{-1}$, $\Omega_m$=0.3, and $\Omega_\Lambda$=0.7 throughout.

Equation \ref{dn} has two unknowns to be explored, the luminosity function term being the more straightforward.  Numerous analyses have explored this topic in various contexts (\cite{schmidt01, lloyd02, norris02, stern02, guetta05,lloyd04, daigne06, coward05, cobb06, pian06, soderberg06, chapman07, liang07, dai09,virgili09, qin10, wanderman10}).  We adopt the generally accepted broken power-law model:
\begin{equation}\label{BPL}
\Phi(L)=\Phi_0\left[\left(\frac{L}{L_b}\right)^{\alpha_1}
+\left(\frac{L}{L_b}\right)^{\alpha_2}\right]^{-1}~
\end{equation}
where $\alpha_1$ and $\alpha_2$ are the power law 
indices, $L_b$ the break luminosity, and $\Phi_0$ a normalization 
constant.  We consider solely `classical' high-luminosity GRBs, ignoring local low-luminosity events and their contribution to the luminosity function (\cite{coward05,liang07,le07,virgili09}) in order to have as unbiased a sample as possible.

The GRB rate, $R_{GRB}(z)$, is the main focus of this analysis as it is a convolution of the star-formation history, metallicity and evolution effects.  Next, we present the specifics of each SFH model and enhancement.

\subsection{Cosmic star-formation history}
The cosmic star-formation history is the basis for the rate distribution from which we choose our redshift values for the simulated GRBs.  Many forms are available in the literature but it is generally believed that the SFH increases rapidly to about $z \sim 1-2$ then slowly falls off toward higher redshift, and we use a variety of forms presented in the literature.  Hopkins and Beacom (2006, or `HB') have compiled a widely accepted model fit from numerous multi-band observations (see \cite{handb2006} and reference therein).  Bromm and Loeb (2006 or `BL') present a comprehensive model for the SFH based on a flat $\Lambda$CDM cosmological model with the added contribution of Population III stars at high redshift.  As a control we also include the SF2 model of Porciani and Madau (2001, or `PM'; See Figure \ref{SFH}) based on estimates from UV-continuum and H$\alpha$ emission. A list of SFH models used are summarised in Table 1.

In addition, we utilize a model derived from cosmological SPH simulations of Choi \& Nagamine (2010, or `CN').
They developed a modified version of GADGET-3 code (originally described in \cite{Springel:05}), including radiative cooling by H, He, and metals \citep{Choi.Nagamine:09}, heating by a uniform UV background of a modified Haart and Madau (1996) spectrum \citep{Katz.etal:96,Dave.etal:99}, a sub-resolution model of multiphase ISM \citep{Springel.Hernquist:03}, the ``Pressure'' star formation model \citep{Schaye.etal:10, choiandnag10}, and the ``Multicomponent Variable Velocity'' galactic wind model (\cite{Choi.Nagamine:11}). They have shown that the metal line cooling enhances star formation across all redshifts by about $10-30$\% \citep{Choi.Nagamine:09}, and that the Pressure SF model suppresses star formation at high-redshift due to a higher threshold density for star-formation \citep{choiandnag10} with respect to the earlier model by \citet{Springel.Hernquist:03}.
\citet{Choi.Nagamine:11} also showed that the MVV wind model, which is based on both momentum-driven and energy-driven galactic winds, makes the faint-end slope of the GSMF slightly shallower compared to the constant velocity galactic wind model of \citet{Springel.Hernquist:03}. The adopted cosmological parameters are consistent with the WMAP best-fit values \citep{Komatsu.etal:11}: $\Omega_m = 0.26$, $\Omega_{\Lambda} = 0.74$, $\Omega_b = 0.044$, $h=0.72$, $n_{s}=0.96$, and $\sigma_{8}=0.80$.  These values are only slightly different from those presented in \S~2.  The results from three simulations with box sizes of comoving $10, 34$ and 100$h^{-1}$\,Mpc were combined to obtain a full SFH, including galaxy stellar masses above $10^7\,M_{\odot}$ \citep{Choi.Nagamine:11b}.

\begin{table}
\centering
\caption{Summary of star formation-history models}
\begin{tabular}{cc}
\hline
\hline
SFH Model & Reference \\
\hline
PM & Porciani \& Madau (2001) \\
HB & Hopkins \& Beacom (2006) \\
BL & Bromm and Loeb (2006) \\
CN & Choi \& Nagamine (2010) \\
\hline
\end{tabular}
\end{table}

\subsection{Metallicity}

One popular explanation for possible enhancements of the GRB rate is the effect due to decreasing metallicity with redshift (\cite{fynbo03,conselice05,gorosabel05,chen05,starling05,langerandnorm06, li08, qin10,butler2010,campisi10}).  If GRBs occur more frequently in low-metallicity environments, then this could be a possible mechanism for enhancing the GRB rate at high redshift.  Langer \& Norman (2006; LN) proposed an analytical form for the mass density fraction in galaxies with a mass less than $M$, based on the galaxy stellar mass function (GSMF).  This mass is then related to the amount of metals via the galaxy mass-metallicity relation (\cite{tremonti04,savaglio05}), giving:
\begin{equation}\label{psi}
\Psi \left(\frac{Z}{Z_\odot}\right) = \frac{\hat{\Gamma}[\alpha_G+2, (Z/{Z_\odot})^\beta 10^{0.15\beta z}]}{\Gamma[\alpha_G+2]},
\end{equation}
where $\hat{\Gamma}$ and $\Gamma$ are the incomplete and complete gamma functions, $\beta$ the power-index of the galaxy mass-metallicity relation, and $\alpha_G$ the faint-end slope of the GSMF.  To begin, we utilize this relation with constant parameters from the literature ($\alpha_G=-1.16$, $\beta=-2$, $\epsilon = (Z/Z_\odot) = 0.1$; \cite{langerandnorm06,li08}) and then build upon it to formulate a weighted version to account for the more realistic case of variations in metallicity from $\epsilon=0.1-0.4$ (Figure \ref{li}).  This function also contains many assumptions about the underlying mass distribution and mass-metallicity relation which we will expand on in $\S 4.4$.

The cosmological simulations derive star formation rates for populations of stars from various metallicities without the need of an external expression. At every time step, star particles are created in high-density regions that exceed the threshold density according to the star-formation law matched to the locally observed Kennicutt (1998) law.  Once a star particle is created, instantaneous recycling is assumed, and the metals are ejected with an yield of $Y=0.02$ and distributed to the nearby environment by a galactic wind. 
Niino et al. (2011) have used similar simulations to examine the metallicity of GRB host galaxies, and found good agreement with observations.  See Figure \ref{li} for a comparison of Equation \ref{psi} and the result from cosmological simulations.

\subsection{Rate evolution}

Motivated by the literature (e.g. \cite{kistler08, kistler09, qin10}) we include a discussion on an increase in the GRB rate with redshift as $(1+z)^\delta$.  This is not to be confused with the evolution of the break luminosity of the luminosity function, which has the same functional form (See \S2.4).  The latter has a more physical meaning (e.g. GRBs becoming brighter with increased redshift), whereas the former is more of a general statement of the GRB rate, increasing in this fashion due to an unspecified process.  The former may be related to an evolving stellar initial mass function with redshift that causes a shift to a top-heavy stellar IMF (\cite{wang11}). These simulations assume $R_{GRB} \propto SFH \times (1+z)^\delta$ with a non-evolving luminosity function. The results of these simulations are presented in \S 4.5.

\subsection{$L_b$ evolution}

The last form of enhancement of the GRB rate is evolution of the break of luminosity function, $L_b$, with redshift.  We take a similar functional form to the rate evolution, assuming that the luminosity function break evolves as $(1+z)^\gamma$.  This increase manifests itself as an increase of bright bursts at higher redshifts, which increases their detection rate.  Unlike the previous section, these simulations assume that $R_{GRB}$ is proportional to the SFH and that the luminosity function break, $L_b$, evolves as $(1+z)^\gamma$.  The results of these simulations are presented in \S 4.6

\subsection{Threshold and other details}

Once a luminosity and redshift pair is chosen according to the distributions discussed above, it is necessary to adopt a threshold condition that mimics the detector in question.  We adopt the threshold condition based on the probability of triggering Swift derived by Qin et al. (2010):
\begin{equation}
\eta_t =  \left \{
\begin{array}{l l}
5.0P^{3.85}, & P<0.45 \\
0.67(1.0-0.40/P)^{0.52}, & P \geq 0.45
\end{array}
\right.
\label{trigger}
\end{equation}
where P is the photon flux of the burst in the 15$-$150 keV band.  This equation is based on the similarities between the BATSE and \textit{Swift} photon flux samples.  By comparing the relative number of bursts, both triggered and un-triggered, that occur in a particular photon flux bin to the total number of bursts it is possible to obtain a probability for triggering that instrument.  Qin et al. (2010) were able to fit the distribution to derive Equation \ref{trigger}.  When comparing different detectors it is necessary to have a normalized value for P, and Qin et al. found that the normalization for both detectors is similar, so one expression is a good approximation for both detectors.  For comparisons with the observed redshift and luminosity samples, we include an additional probability for the detection of a redshift since not all bursts have redshifts.  Similarly, a probability of assigning a redshift is found by looking at the distribution of bursts with redshift versus the total number of bursts per redshift bin.  Qin et al. (2010) did not find a large distinction between these two samples but parametrize the probability as
\begin{equation}
\eta_z=0.26+0.032e^{1.61 \log P}.
\end{equation}
Both expressions and the $\log N - \log P$ analysis depend on the calculation of the photon flux.  The \textit{energy} flux is calculated directly from $F_{pk}=L/4 \pi D^2_L(z) k$ where $D_L(z) $ is the luminosity distance at a given redshift, and $k$ is the $k$-correction
\begin{equation}\label{kcorr}
k=\frac{\int_{1/(1+z)}^{10^4/(1+z)}EN(E)dE}{\int_{e_1}^{e_2}EN(E)dE},
\end{equation}
which corrects the flux from the bolometric $1-10^4$ keV band into the observed band $(e_1,e_2)$.  In Equation \ref{kcorr}, N(E) is the photon spectrum of the GRB, which we assume to be a Band function (\cite{band93}).  The Band function is a smoothly joined power law function that has pre- and post-break slopes $\alpha$ and $\beta$ around a break energy $E_0$.  The peak of the $\nu F_\nu$ spectrum, $E_{peak}$, is related to this energy by $E_{peak}=(2+\alpha)E_0$.  Since the peak energy of bursts changes with the energy of the burst (i.e. the Amati relation; \cite{amati02, liangdai04}) we utilize the relation derived by Liang et al. (2004) to assign values for the simulated $E_{peak}$:
\begin{equation}\label{EpLiso}
 E_{peak}/200 {\rm keV}=C^{-1} (L/10^{52} {\rm erg\ s}^{-1})^{1/2},
\end{equation}
where $C$ is a random uniform deviate between [0.1,1] and $L$ the luminosity.  We also randomly sample $\alpha$ and $\beta$ uniformly between $-0.83 < \alpha < -1.2$ and $ -2.1 < \beta < -2.5$, which roughly correspond to the observed limits of these values. From these spectral parameters and peak energy flux, F, we then calculate the peak photon flux in the detector energy band $(e_1,e_2)$ via
\begin{equation}\label{Photon}
P=\frac{F\int_{e_1}^{e_2}N(E)dE}{\int_{e_1}^{e_2}EN(E)dE}.
\end{equation}

\section{Testing for consistency}

Once we have a set of simulated bursts that are `detected' and follow the GRB rate and luminosity function described above, it is necessary to test the consistency with the observed data.  Our sample consists of 166 \textit{Swift} and \textit{HETE} era GRBs with known redshift through September 2009.  We remove the Type I GRBs (see \cite{zhang09,virgili11,kann11}), outlying low-luminosity bursts (GRB 980425, 060218), and any bursts with disputed or non-secure redshifts. This assumes that the observed redshift sample is the true intrinsic sample, although the detection of redshifts depends on a variety of observational factors and potential biases (\cite{fiore07,jakobsson11}).  In order to calculate the bolometric peak luminosity it is necessary to have the energy or photon flux as well as the spectral information for the $k$-correction.  Most \textit{Swift} bursts are fit by a simple power-law spectrum due to small band pass of the \textit{Swift} detector (\cite{sakamoto07}).  We instead assume all bursts to have a Band function spectrum (\cite{band93}) utilizing the observed pre-break slope and assuming the typical value of $2.5$ for the post-break slope, as assuming a simple power-law extending to high energies will inevitably overestimate the high-energy contribution to the flux.  This assumption is validated by observations of GRBs observed jointly with \textit{Fermi} LAT and GBM, where a Band function spectrum is seen over many orders of magnitude and extending to high energies (\cite{abdo09,zhang11}).  $E_{peak}$ is used from the literature unless absent, in which case the value from catalogue of Butler et al. (2010) is used.   

Next, the simulated set of 175 simulated bursts is compared to the observational sample with the k-sample Anderson-Darling (AD) test for consistency between two distributions (\cite{scholz87}), giving the first three constraints ($L$ alone, $z$ alone, $L-z$ together) to our models.  Each criterion gives a contour showing the consistency with the observed sample in the $(\alpha_1, L_b)$ plane (Figure \ref{hbliw}a) with $\alpha_2$ a constant of 2.2 or 2.5.  We test both values of $\alpha_2$ and results indicate which slope was used. Results are generally insensitive to the choice of $\alpha_2$ and we take values quoted in the literature (\cite{liang07,virgili09,qin10}). 

The two $\log N-\log P$ tests are also conducted with the AD test and compares the simulations to the \textit{CGRO}/BATSE and \textit{Swift} photon flux samples.  In order to have the most unbiased and complete sample from each, they are truncated at $0.4$ [50-300 keV] and $1 \rm~ph~cm^{-2}~s^{-1}$ [15$-$150 keV], respectively (see \cite{landw,band06}) and compared to the 1143 BATSE and 380 triggered \textit{Swift} bursts.  A summary of various models and statistical results are presented in Tables 2-4.

\section{Results}

\subsection{GRB rate $\propto$ SFH}

Clearly the simplest scenario possible for the GRB rate, this set of simulations showed little consistency with the current observations.  Out of the four possible SFH models (Porciani \& Madau (2001), Hopkins and Beacom (2006), Bromm and Loeb (2006), Choi and Nagamine (2010)), only the Bromm and Loeb model showed consistency with the observations (Table 2).  The luminosity function is generally constrained to be shallow, with pre-break slopes generally more shallow than $\sim -0.2$, and shows that there is the need for some form of increase of the rate compared to other SFH models at higher redshifts.  Determining the form and possible cause(s) of this increase in GRB rate is a major goal of this analysis.  This model, however, is based on the theory that the rate enhancements produced at high $z$ are attributed to the contribution of Population III stars that were developing in the early universe  $z \sim~15$.  To date, observations have not shown that GRBs arise from Population III stars (\cite{tanvir09,salvaterra09}) and we caution drawing an association based solely on the form of this SFH.  

\begin{figure}
\includegraphics[scale=0.8]{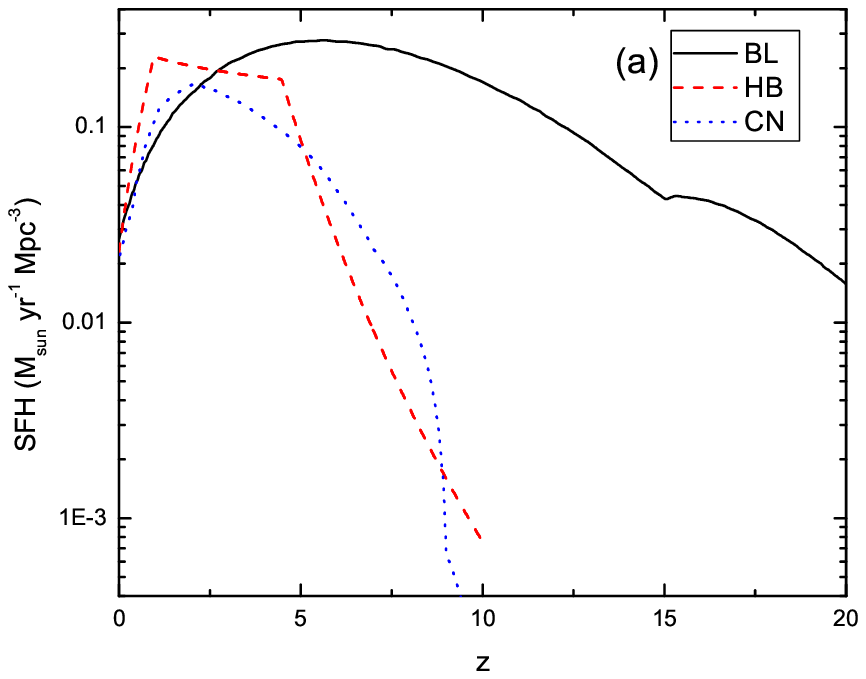}
\includegraphics[scale=0.8]{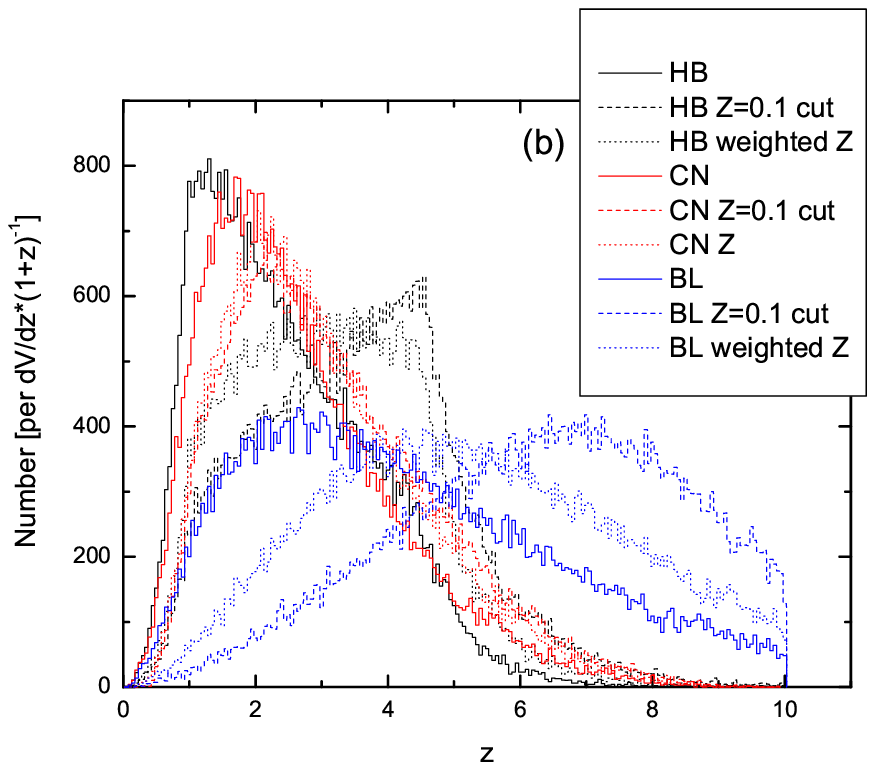}
 \caption{Panel (a): Star-formation history models utilized in the analysis. Panel (b): Simulated relative number of GRBs per unit co-moving volume$\times(1+z)^{-1}$ for different models.  This panel shows the output from the code without a threshold, so as to check the underlying distribution and see the relative affects of the metallicity relations on the base SFH.}
 \label{SFH}
\end{figure}

\subsection{GRB rate $\propto$ SFH+Metallicity cutoff}

The next step, as detailed above, is to consider the addition of a term that in some manner accounts for an increase in the GRB rate in lower metallicity environments.  A key component believed to aid in the creation of Type II GRB is a fast-rotating core of the progenitor star.  Low metallicities may help in reducing the mass-loss rate and retain sufficient angular momentum to keep the star rotating quickly and assist in the formation of the GRB jet.  First we consider the formalism of Langer and Norman (2006; see also \cite{li08,qin10}) as detailed in Eqn \ref{psi}, for all models with the exception of CN, as they form stars self-consistently according to metal line cooling rates in the simulation.  The derivation is straightforward and clear in Langer and Norman (2006), however, there are various assumptions in this model that need to be addressed.  The basis for this relation is the \textit{galaxy} stellar mass function (GSMF), which is assumed to be a Schechter function (i.e. a power law with exponential cutoff; Eqn 1 in \cite{langerandnorm06} and references within).  The amount of galaxy stellar mass within a mass $M$ is then related to the amount of metals by the mass-metallicity relation, of the form $M/M^* = K(Z/Z_\odot)^\beta$, where $K$ and $\beta$ are constants that are constrained by observation (\cite{tremonti04, savaglio05}) and $M^*$ is the characteristic mass of the GSMF.  Previous studies do not address the scatter and/or evolution with redshift of the GSMF faint-end slope $\alpha_G$,
and assume that the average cosmic metallicity of the Universe decreases as $d[Z]/dz = -0.15$ dex.  The normalization of the mass-metallicity relation changes with redshift and for a particular metal cut $\epsilon = (Z/Z_\odot)$, the mass fraction of metals also changes.  This effect is reflected in the $10^{0.15 \beta z}$ term of Equation \ref{psi}. Changes to this term are not examined explicitly.  

Using this relation with the parameters assumed in Langer and Norman (2006) ($\alpha_G=-1.16,~\beta=2$) and a cutoff metallicity for production of GRBs of $\epsilon = Z/Z_\odot=0.1$, we find that no models agree with the $L$ and $z$ constraints to the 2$\sigma$ level.  The cosmological simulation results are similar and show that a strict metal cut at $Z/Z_\odot=0.1$ is insufficient to explain the observations.

\begin{figure}
\includegraphics[scale=0.8]{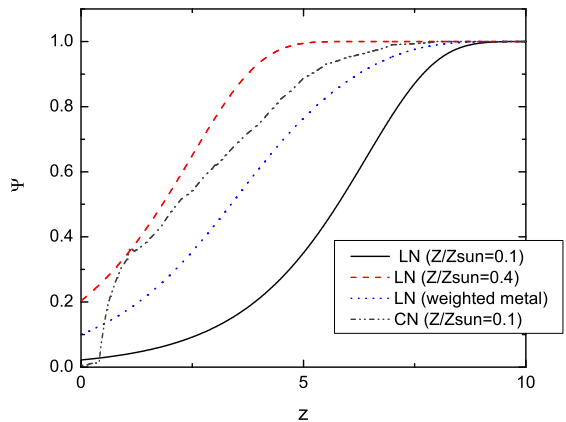}
 \caption{Fractional stellar mass density contained in galaxies with metallicities below $Z/Z_\odot$ (Eq.\,\ref{psi}) from Langer and Norman (2006, LN), including different metal cuts ($Z/Z_\odot = 0.1, 0.4$) and modifications from weighting.  The expression derived from the star-formation history of CN with a metal cut of 0.1 is included for comparison.}
 \label{li}
\end{figure}

\subsection{GRB rate $\propto$ SFH+Weighted metallicity}

Building upon the previous section, we introduce a weighting to the metallicity cut in order to broaden the scope of Eqn \ref{psi}.  It is more realistic to consider a range of metallicities in which GRBs can occur, especially since GRBs have been observed in environments with metallicity greater than 0.1 (\cite{holland10,levesque10}).  Instead of taking the value of the relation from Langer and Norman (2006) for a particular value of $\epsilon$, we instead weight the effect of different metallicities, ranging from $Z/Z_\odot=0.1-0.4$, for a particular redshift.  We have utilized a Gaussian (with $\sigma=0.1$ or $0.2$) to weight the contributions of metals so that there is an exponential (rather than sharp) cut above the critical metallicity.  The contributions from various metallicities are then added with proper weighting to produce an `effective' $\Psi$ (Equation \ref{psi}). This approach yields an intermediate solution between strict metal cuts of 0.1 and 0.4 (Figure \ref{li}).  A similar approach is taken for the CN model utilizing the SFR for different metal cuts provided from the simulation instead of applying Eqn \ref{psi}.

Using this formulation we re-run the previous SFH models and find that the HB model is the only model that can pass all of the observational tests, including the $\log N-\log P$, giving luminosity function parameters in the range of $(\alpha_1, \alpha_2=2.2, L_b)=(0.11-0.19, 2.2, 6-10\times10^{52}~\rm erg~s^{-1})$ (Figure \ref{hbliw}, Table 2).   The BL model, with its intrinsically large rate at high-$z$, overproduces bursts at high-$z$ when metallicity is added.

\subsection{More on the metallicity approximation}

\begin{figure}
\includegraphics[scale=0.8]{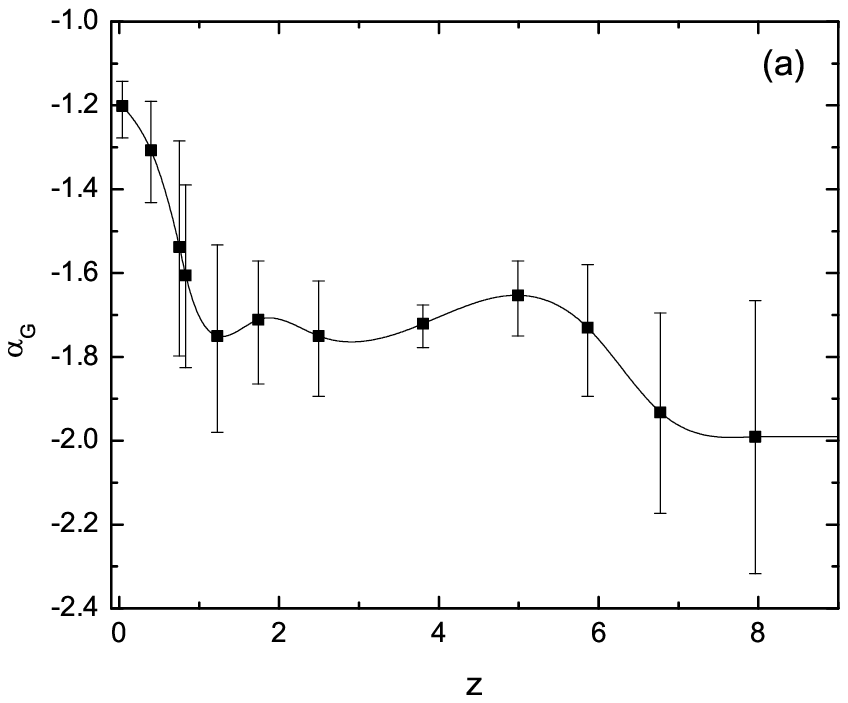}
\includegraphics[scale=0.8]{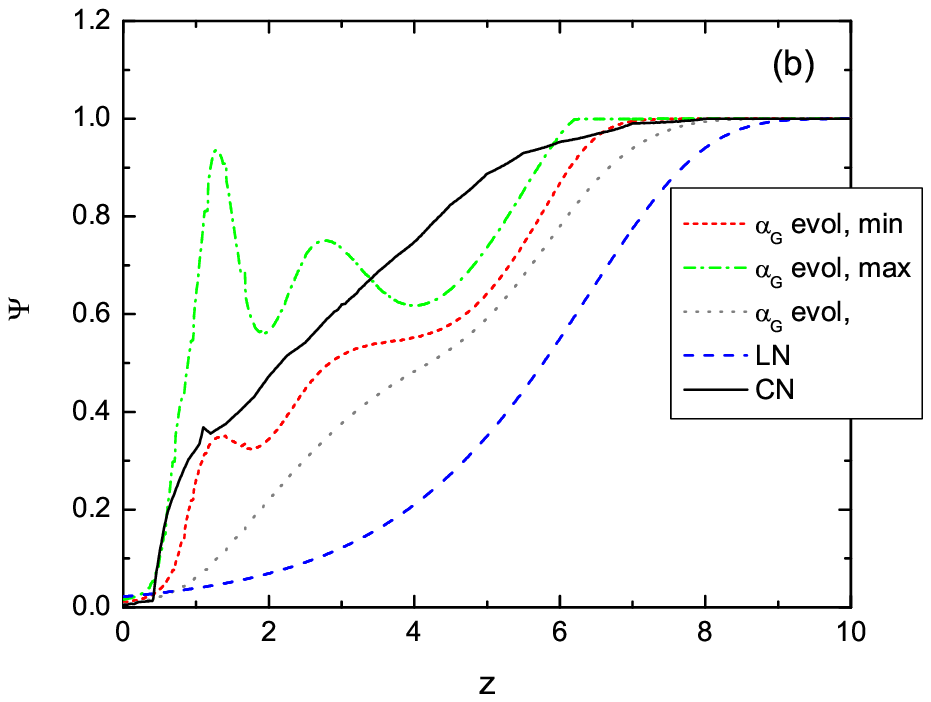}
 \caption{Panel (a): Redshift evolution of the GSMF faint-end slope, $\alpha_G$, including error bars (Bouwens et al. 2010). Panel (b:) Effect of the evolution of $\alpha_G$ on the expression from LN.  The $Z/Z_\odot=0.1$ cut expression from LN and CN are included for comparison.}
 \label{alpha}
\end{figure}

\begin{figure*} 
\centerline{\includegraphics[width=9.2cm]{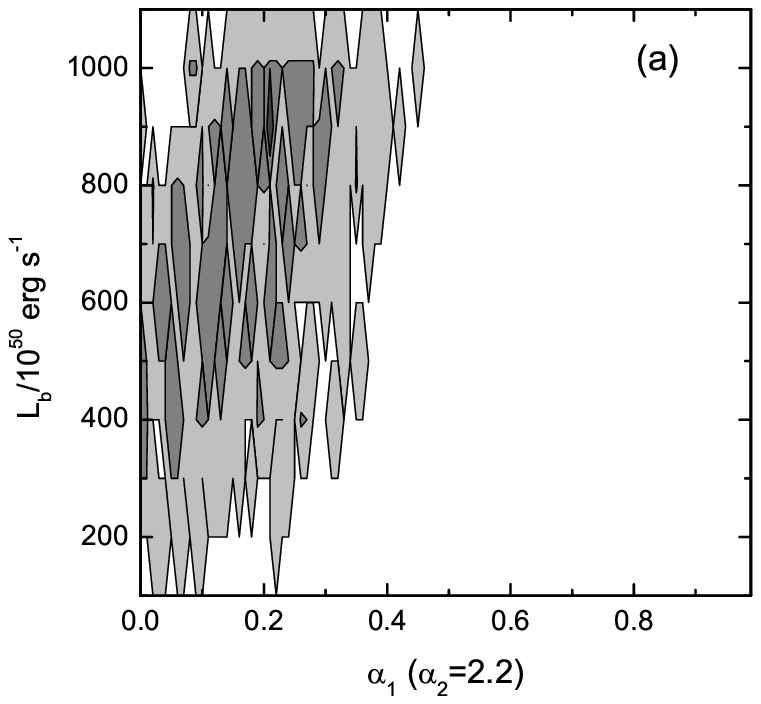} \includegraphics[width=8.8cm]{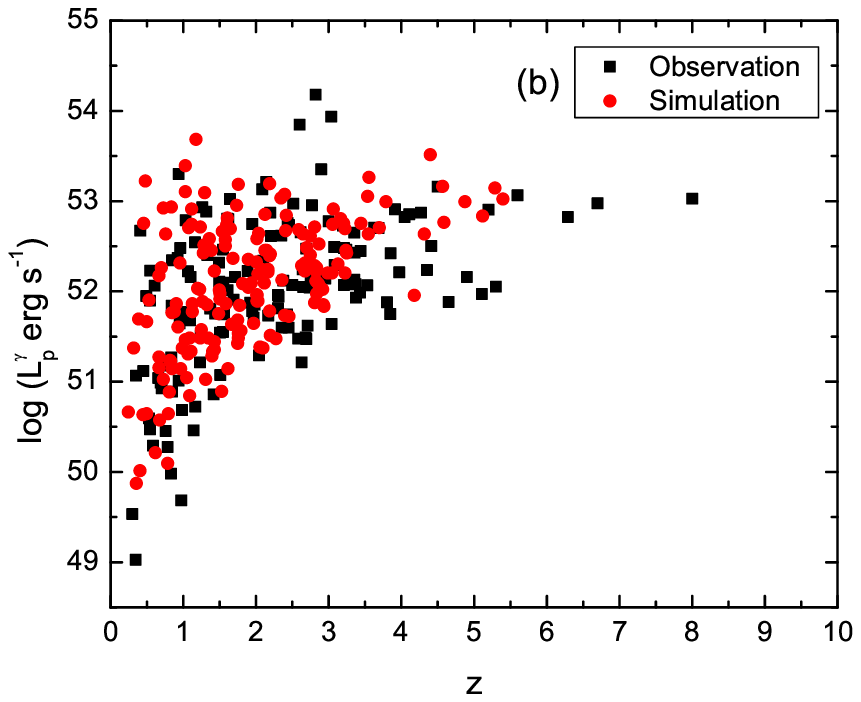}} 
\centerline{\includegraphics[width=9.2cm]{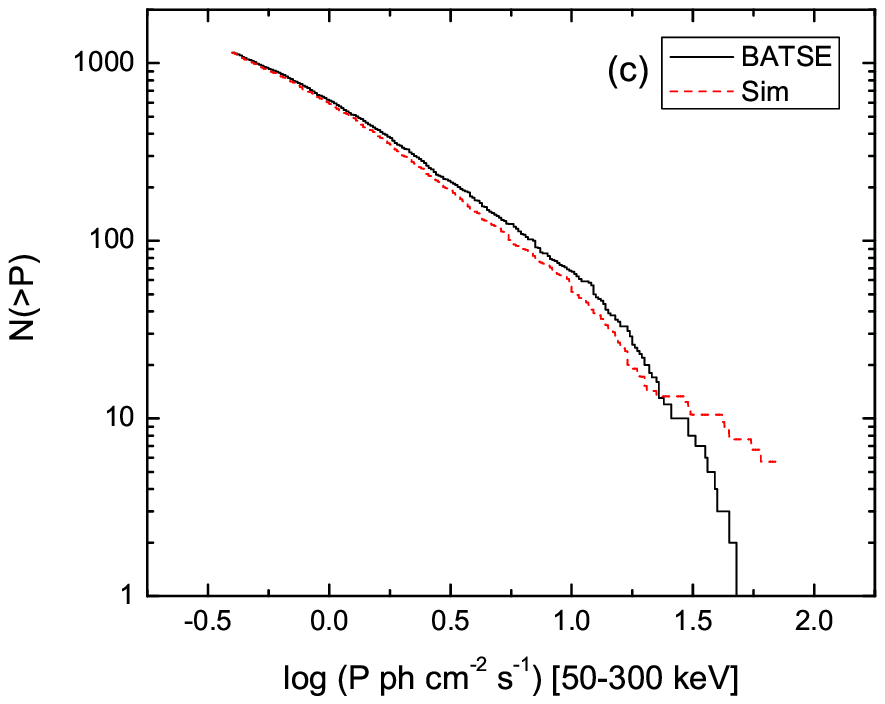} \includegraphics[width=8.8cm]{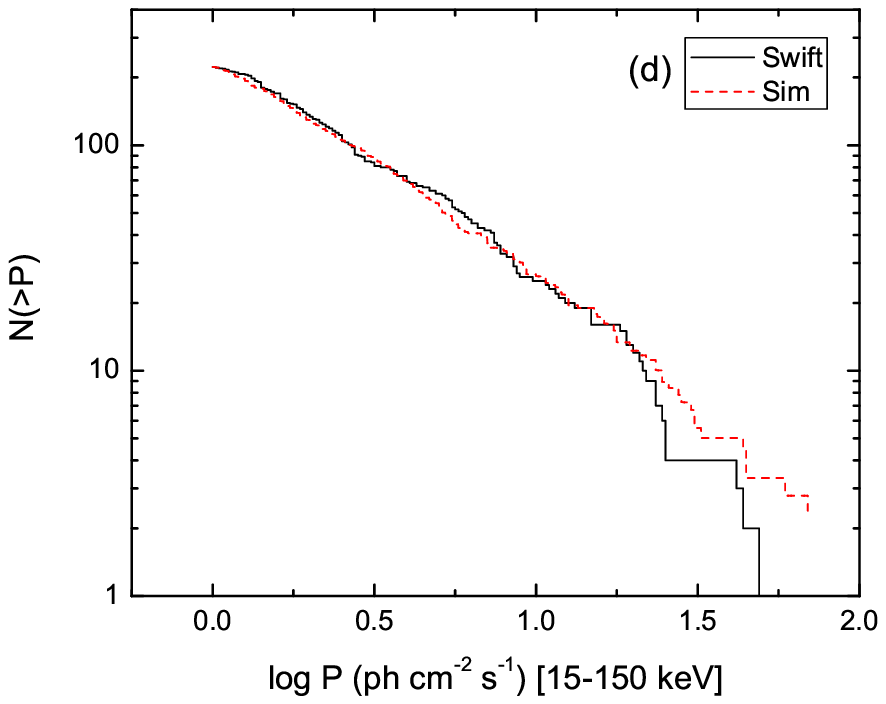}} 
\caption{HB SFH model with the weighted LN expression. Panel (a): Contours for consistency in BOTH $L$ and $z$.  Dark grey = 2$\sigma$ contour and light grey 3$\sigma$ contour. Panel (b): Sample 2D distribution from area of consistency in 2$\sigma$ region, $(\alpha_1, \alpha_2, L_b)=(0.15, 2.2, 8\times 10^{52} \rm~erg~s^{-1})$.  Panels (c) and (d): BATSE and \textit{Swift} $\log N-\log P$ for same parameters as panel (b).}
\label{hbliw}
\end{figure*}

Up to this point, the analysis does not directly compare how the assumed metallicity relation (Eqn \ref{psi}) affects the GRB rate compared to the cosmological simulations.  This is an important and related topic, since the models that use the LN expression show consistency with the observations of Type II GRBs, while the more rigorous and complete method of CN to calculate the metallicity shows no consistency.  The differences, we come to find, are non-negligible.  Why would the HB and CN models, whose total star-formation rates are quite similar, differ so largely when their respective metal cuts are applied (Figure \ref{SFH}b)?  The relation from Langer and Norman (2006) is an approximation to a very complex problem in galaxy evolution.  The cosmological simulations by Choi and Nagamine (2010) address a variety of effects that contribute to the metal distribution (such as mixing due to galactic outflow and tidal disruption), and calculate the star-formation rate selfconsistently according to varying metal line cooling rate. From those values a realistic view of how the total rate is affected by the reduction in metallicity can be calculated, which is just what Eqn \ref{psi} shows: the net effect to the total star-formation rate by a metallicity cut at $Z/Z_\odot$. The curves for various values of $Z/Z_\odot$ are shown together with the equivalent expression from the CN (Figure \ref{li} and \ref{alpha}).  

Since these expressions are so different, we attempt to look at the structure of the Langer and Norman (2006) expression and see if any part(s) can be improved to create a more realistic view of how metallicity affects the rate of GRBs.  The first major assumption in Eq.~\ref{psi} is the constant value of the GSMF slope, $\alpha_G$, which is observed to be steepening with $z$ (\cite{bouwens10} and references therein), suggesting a larger number of lower-luminosity galaxies at higher redshifts.

Bouwens et al. (2010) detailed several observations of galaxies at $z \sim 7-8$ and summarised the evolution of the luminosity function of galaxies.  From their Figure 15 we are able to extract the slope of the GSMF as a function of redshift to incorporate into our code.  Using a spline fit and cubic interpolation we are able to approximate the behavior of $\alpha_G$ both at the data point as well as maximum and minimum values from the error bars provided, which range from about $-1 > \alpha_G > -2$ in the range $z \sim 0-10$ (Figure \ref{alpha}).  We consider only values above $\alpha_G = -2$ as the metallicity relation is undefined at the value $\alpha_G+2=0$, which affects the minimum error bar approximation.  For that case we assume $\alpha_G$ is constant with the value of the lowest data point (-1.99) out at higher redshifts.  Above $z=8$ we again assume all values of $\alpha_G$ are constant.  As shown in Figure \ref{alpha}, the evolution of $\alpha_G$ implies a faster cosmic metallicity enrichment than just applying the unaltered expression and pushes the curve toward lower redshift and closer to the results from cosmological simulations.  Using the values of the upper error bars of Figure \ref{alpha}a gives solutions that are similar to the weighted expression of LN (Table 3).

This more realistic approach show some consistency with observation, but only in a few cases.  The HB model utilizing the upper limits of the evolution of $\alpha_G$ and a metal cut of $Z/Z_\odot=0.1$ shows some consistency with the $L$ and $z$ constraints and consistency with all $\log N-\log P$ constraints, while models with metallicity weighting show results only in the 3$\sigma$ contour.  All other HB models show little consistency in all tests while the BL model shows some areas in the $L$ and $z$ constraints but shows a large deviation in the $\log N-\log P$ results for both BATSE and BAT. The lack of consistency that is evident in most models is generally attributed to an overproduction of bursts at $z \sim 1-2$.  Models utilizing the original framework of LN may show more consistency with the observations, but these set of updates are a promising and necessary direction for study that, with further enhancements, might be able to fully explain the rate increase.

\subsection{GRB rate $\propto$ SFH $\times (1+z)^\delta$}

As detailed in the literature (\cite{kistler08,kistler09,qin10}) we also consider an increase in GRB rate as $(1+z)^\delta$, where $\delta= 0.2,0.5,0.8$.  We consider all of the SFH models with and without metallicity enhancements (no GSMF evolution) and find that a few of these models are able to pass the $L$ and $z$ constraints but fail to pass the BATSE and \textit{Swift} $\log N-\log P$ constraints (Table 4).  

\subsection{Luminosity function break evolution $\propto (1+z)^\gamma$}

\begin{figure*} 
\centerline{\includegraphics[width=9.2cm]{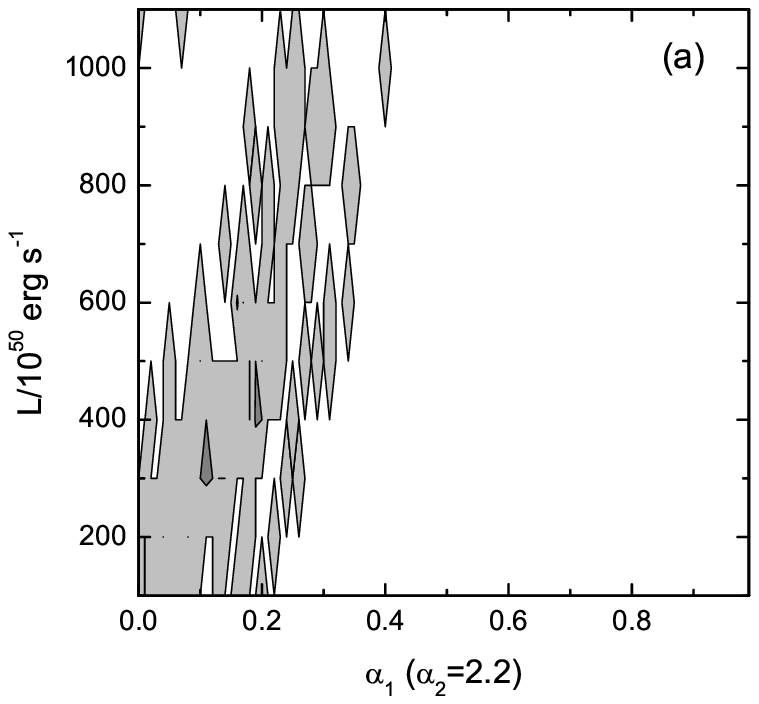} \includegraphics[width=8.8cm]{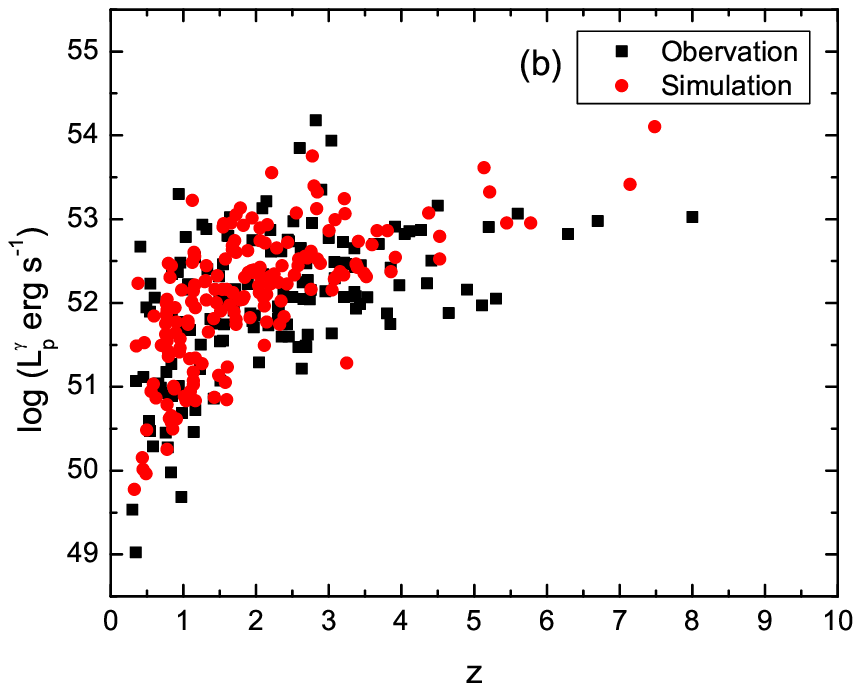}} 
\centerline{\includegraphics[width=9.2cm]{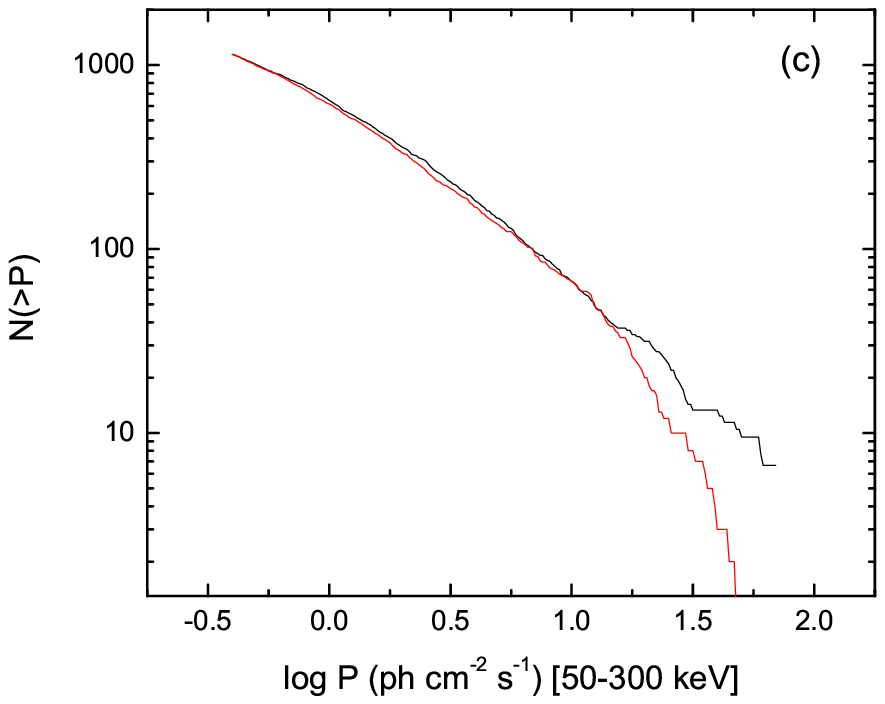} \includegraphics[width=8.8cm]{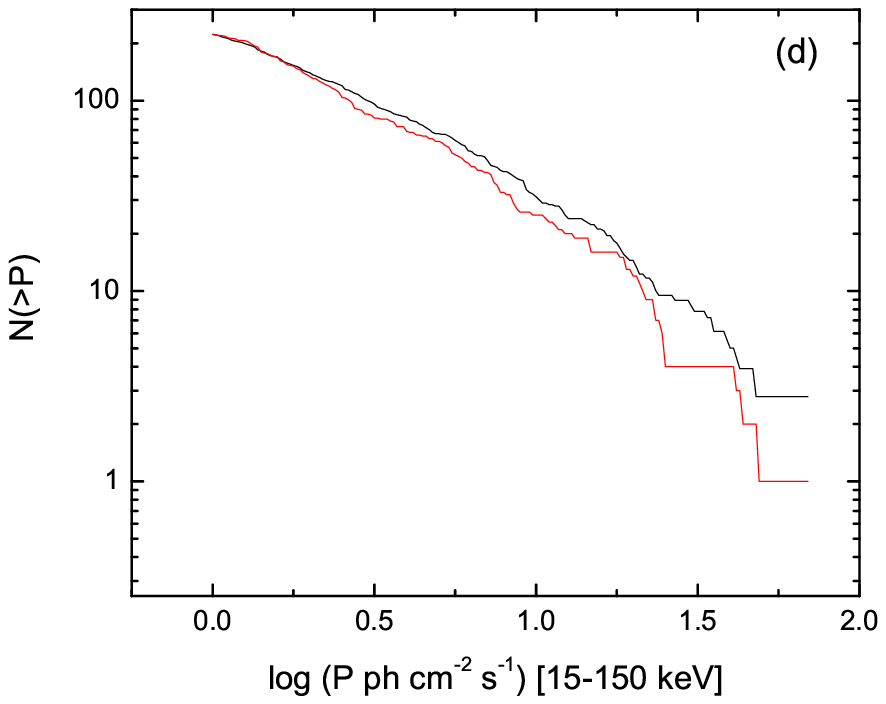}} 
\caption{CN SFH model with luminosity break evolution $\propto~(1+z)^{1.0}$. Panel (a): Contours for consistency in BOTH $L$ and $z$.  Dark grey = 2$\sigma$ contour and light grey 3$\sigma$ contour. Panel (b): Sample 2D distribution from area of consistency in 2$\sigma$ region, $(\alpha_1, \alpha_2, L_b,\gamma)=(0.05, 2.2, 3\times 10^{52} \rm~erg~s^{-1}, 1.0)$.  Panel (c) and (d): BATSE and \textit{Swift} $\log N-\log P$ for same parameters as panel (b).}
\label{cosmo}
\end{figure*}

Lastly we consider the evolution of the luminosity function break as detailed above.  We see some consistency with the CN model and evolution with $\gamma \sim 0.5-1.5$.  The $3\sigma$ regions for the CN models show areas of consistency, with a few showing $2\sigma$ significance (i.e. $\gamma=1.0, 1.3$) (Figure \ref{cosmo}).  The general trend is again for shallow luminosity function slopes, the best models occurring in the area of $(\alpha_1, \alpha_2, L_b,\gamma)=(0.5,2.2,3\times 10^{52} \rm ~erg~s^{-1}$, 1.0). The HB models show some consistency to 3$\sigma$ in the same regions, although not as broadly as the CN model (Table 4).
	
\section{Summary and discussion}

Our work supports the idea that the GRB rate is enhanced at higher redshift (\cite{daigne06,le07, guetta07, li08, kistler08, kistler09, salvaterra09,salvaterra09b,salvaterra07, qin10, wanderman10}). The form of this increase, however, is still unclear.  We have tested various SFH models and enhancements to the GRB rate, reflecting possible effects from changing cosmic metallicity and other evolutionary effects, with a Monte Carlo code.  The resulting output was then tested for consistency with a variety of available \textit{Swift} and BATSE data, including the $L$, $z$, and peak photon flux distributions.  Even when considering a numerical simulation model that takes into account a variety of realistic galactic evolution effects, both with and without metal cuts, and a metallicity relation based on the GSMF (\cite{langerandnorm06}) our models do not show strong consistency with the observed sample, although we believe this is the right direction for this type of study.  This may indicate that metallicity is not solely responsible for the increased rate and that perhaps some other type of enhancement is needed.  To this end, we test both GRB rate evolution and luminosity function (break luminosity) evolution with redshift, finding that the latter is allowed within the constraints of the BATSE and \textit{Swift} data with moderate ($\propto L_b\times(1+z)^{\sim 0.8-1.2}$) evolution.  This statement has, of course, a few caveats.  Embedded in the metallicity relation are a variety of assumptions about the GSMF and the observed mass-metallicity relation.  Laskar et al. (2011) show, using HST observations of GRB host galaxies, that the metallicity relationship likely evolves between redshifts of 3$-$5, which would further affect the results. It is possible that other combinations of parameters or assumptions might yield a more realistic relation, and we suggest further work on how the GSMF and stellar IMF work in tandem to affect the problem at hand.  In addition, recent works have studied the $M-Z$ relation of Type II GRBs and found that the hosts lie below the SDSS $M-Z$ relation (\cite{kocevski11,mannucci11,campisi11}).  This adds further evidence to the fact that the assumption of this relation for these types of bursts is likely not valid, and perhaps a consequence of the active star-formation environment instead of a strict metallicity cut.  We have explored some basic changes, such as the evolution of the GSMF faint-end slope, but a comprehensive study of this relation or a realistic alternative, are needed. 

We have detailed a numerical and statistical approach aimed at understanding the properties of the GRB rate in the context of the cosmic star-formation history, including the constraints from newly discovered high-$z$ bursts and the possibly effects of metallicity and various types of evolution.  Recent works have addressed this problem in similar (\cite{qin10}) and fully analytical (\cite{wanderman10}) ways, and share some common points, although both call on fairly strong evolution of the GRB rate ($(1+z)^\delta \sim 0.6-2)$ which we do not find.  Our work also benefits from the inclusion of a fully numerical star-formation history model (\cite{choiandnag10}) as well as a probing of the metallicity relation and cosmological considerations that may affect the GRB rate which are not included in contemporary works on the subject.  Butler et al. (2010) do not find evidence of strong luminosity function or GRB rate evolution and find that a smoothed metallicity cut of $Z/Z_\odot = 0.2-0.5$, following the metallicity considerations of LN, can account for the observations of the current \textit{Swift} sample, although they acknowledge that there are large errors bars.  They also do not include evolution of the GSMF, which may account for the differences with this work.  In addition, we analyse most components separately, and it is possible that the observed distribution is a superposition of a variety of effects.  With enough computational time the various combinations of effects can and should be tested.

By fitting the redshift distribution and $\log N-\log P$ distribution of BATSE and \textit{Swift} bursts, Campisi  et al. (2010) have reached the similar conclusion that Type II GRBs are unbiased tracers of the star-formation history.  Their analysis supports two possible scenarios: (i) a model with no metal cuts and a strongly evolving luminosity function or (ii) a non-evolving luminosity function with a metal cut of $Z/Z_\odot < 0.3$.  Both scenarios assume and fit a Schechter 
luminosity function.  This results are similar to the results presented here, although the luminosity evolution is stronger for the non-metal cut case and the authors claim such large changes in GRB properties with redshift as unrealistic, favoring a model with a metal cut and no luminosity function evolution.

\section*{Acknowledgments}
This work is supported by NSF through grant
AST-0908362, and by NASA through grants NNX10AD48G and NNX10AP53G.
KN is supported in part by the NSF grant AST-0807491, 
NASA grant HST-AR-12143-01-A, National Aeronautics and Space Administration under Grant/Cooperative Agreement No. NNX08AE57A issued by the Nevada NASA EPSCoR program, and the President's Infrastructure Award from UNLV. This research is also supported by the NSF through the TeraGrid resources provided by the Texas Advanced Computing Center.

\begin{table*}
\scriptsize
\centering
\caption{SFH models and test statistics for a variety of simulations, including a metallicity cut of $Z/Z_\odot=0.1$.  If consistency is found with the $L$ and $z$ samples, indicated by a `Y' in the second column, then the LF parameters are listed with the outcomes of the $\log N - \log P$ analysis.  Later models include the addition of metallicity in the form of the expression from LN, but with a $Z/Z_\odot=0.1$ cut as well as the weighted expression. An `N' in the significance column indicates that test fails beyond a $3\sigma$ level. A similar analysis, with similar results, was conducted for models with the luminosity function post-break slope $\alpha_2=2.5$. Columns 8 and 10 indicate the significance level for the BATSE ad \textit{Swift} $\log N-\log P$ tests, respectively.}
\begin{tabular}{cccccccccc}
\hline
\hline

 Model & L-z? & LF parameters & z &  L & Significance & BATSE LNLP & Sig & \textit{Swift} LNLP & Sig  \\
  & Y/N & $(\alpha_1,L_b, \alpha_2)$ & Stat, P-value & Stat, P-value &  z/L & T stat, P-value &  & T stat, P-value & \\
  \hline
  \textbf{GRB rate $\propto$ SFH} &  \\ 
  $\alpha_2$ \textbf{=2.2} & \\
  HB & N & - & - & - & - & - & - & - & - \\
  BL & Y & (0.01,500,2.2) & -0.63727	, 0.51955 & 1.49613, 0.07869 & 1$\sigma$/2$\sigma$ & 0.92194, 0.13939& 2$\sigma$ & 0.18829	0.28273 & 2$\sigma$ \\
        & Y & (0.11,600, 2.2) & 1.16638, 0.10932 & -0.34006, 0.42921 & 2$\sigma$/1$\sigma$ & 0.42551, 0.22537 & 2$\sigma$ & 0.97859, 0.13178 &  2$\sigma$ \\
        & Y & (0.2,900,2.2) & 1.4281, 0.0842 & -0.33604, 0.428 & 2$\sigma$/1$\sigma$ & 3.72049, 0.01036 & 3$\sigma$ & 0.95844	0.13444 &  2$\sigma$ \\
 CN & N & - & - & - & - & - & - & - & - \\
 PM & N & - & - & - & - & - & - & - & - \\
\hline

 \textbf{GRB rate $\propto$ SFH + Metallicity} & \\
  $\alpha_2$ \textbf{=2.2} & \\

 HB+Li & N & - & - & - & - & - & - & - & - \\
 BL+Li & N & - & - & - & - & - & - & - & - \\
 CN 0.1 cut & N & - & - & - & - & - & - & - & - \\
 HB+Li weighted & Y & (0.11,600,2.2) & 0.88134, 0.14509 & -0.08952, 0.35632 & 2$\sigma$/1$\sigma$ & 1.973, 0.04921& 3$\sigma$ & 0.17974, 0.28485 & 2$\sigma$ \\
          & Y & (0.19,1000,2.2) & 0.24896, 0.26793 & -0.80114, 0.5692 & 2$\sigma$/1$\sigma$ & 0.3929, 0.23235& 2$\sigma$ & 0.48579, 0.21293 & 2$\sigma$ \\
          & Y & (0.15,800,2.2) & 0.29743, 0.25647 & -0.4227, 0.45411 & 2$\sigma$/1$\sigma$ & 1.07391, 0.11987 & 2$\sigma$ & -0.21968, 0.39359 & 1$\sigma$ \\
 BL+Li weighted & N & - & - & - & - & - & - & - & - \\
 CN weighted & N & - & - & - & - & - & - & - & - \\
 \hline

 \hline
\end{tabular}
\end{table*}

\begin{table*}
\scriptsize
\centering
\caption{SFH models and test statistics for models with evolving GSMF faint-end slope, $\alpha_G$, in the metallicity equation of LN.  If consistency is found with the $L$ and $z$ samples, indicated by a `Y' in the second column, then the LF parameters are listed with the outcomes of the $\log N - \log P$ analysis.  `sml scatter' and `lrg scatter' indicate the amount of scatter in the weighting of the metallicity relation, 0.1 and 0.2 respectively.  `central values' and `upper limits' indicate what set of $\alpha_G$ values were used in the analysis, those corresponding to the data point value or the upper limits in Figure 3a. An `N' in the significance column indicates that test fails beyond a $3\sigma$ level.  All models assume the post-break slope of the luminosity function $\alpha_2 = 2.2$. Columns 8 and 10 indicate the significance level for the BATSE ad \textit{Swift} $\log N-\log P$ tests, respectively.} 
\begin{tabular}{cccccccccc}
\hline
\hline

 Model & L-z? & LF parameters & z &  L & Significance & BATSE LNLP & Sig & \textit{Swift} LNLP & Sig  \\
  & Y/N & $(\alpha_1,L_b, \alpha_2)$ & Stat, P-value & Stat, P-value &  z/L & T stat, P-value &  & T stat, P-value & \\
  \hline

 \textbf{Models including $\alpha_G$ evolution} & \\
 \textbf{(data point values)} & \\

 HB+Li+$\alpha_G$ evol & Y & (0.1,800,2.2) & 2.85189, 0.02164 & -0.61141, 0.51165 & 3$\sigma$/1$\sigma$ & 0.21195, 0.2769 & 2$\sigma$ & 0.84384, 0.15056  & 2$\sigma$ \\
 HB+Li+$\alpha_G$ evol+weighting & \\
 (sml scatter) & N & - & - & - & - & - & - & - & - \\
 HB+Li+$\alpha_G$ evol+weighting & \\
 (lrg scatter) & N & - & - & - & - & - & - & - & - \\
 BL+Li+$\alpha_G$ evol & Y & (0.41,900,2.2) & 0.19157, 0.28191 & -0.37904, 0.44092 & 2$\sigma$/1$\sigma$ & 18.10581, 0 & N & 1.49981, 0.0784 & 2$\sigma$ \\
 BL+Li+$\alpha_G$ evol+weighting & \\
 (sml scatter) & Y & (0.39,800,2.2) & 1.6878, 0.06507 & -0.07309, 0.35173 & 2$\sigma$/1$\sigma$ & 35.23425, 0 & N & 2.73904, 0.02395 & 3$\sigma$\\
 BL+Li+$\alpha_G$ evol+weighting & \\
 (lrg scatter) & Y & (0.39,900,2.2) & 0.84106, 0.15097 & -0.70235, 0.53937 & 2$\sigma$/1$\sigma$ & & & 0.74228, 0.16633 & 2$\sigma$ \\
 \hline

 \textbf{Models including $\alpha_G$ evolution} & \\
  \textbf{(upper limit)} & \\
 HB+Li+$\alpha_G$ evol & Y & (0.05,600,2.2) & 2.5823, 0.02763 & -0.17585, 0.38088 & 2$\sigma$/1$\sigma$ & 4.42284, 0.00585 & 3$\sigma$ & 0.04375, 0.31989 & 2$\sigma$ \\
 HB+Li+$\alpha_G$ evol+weighting & \\
 (sml scatter) & N & - & - & - & - & - & - & - & - \\
 HB+Li+$\alpha_G$ evol+weighting & \\
 (lrg scatter) & N & - & - & - & - & - & - & - & - \\
 BL+LN+$\alpha_G$ evol & N & - & - & - & - & - & - & - & -\\
 BL+LN+$\alpha_G$ evol+weighting & \\
 (sml scatter) &N & - & - & - & - & - & - & - & - \\
 BL+LN+$\alpha_G$ evol+weighting & \\
 (lrg scatter) & N & - & - & - & - & - & - & - & -\\
 
 \hline
\end{tabular}
\end{table*}

\begin{table*}
\scriptsize
\centering
\caption{SFH models and test statistics for models with rate evolution proportional to $(1+z)^\delta$ and luminosity function break luminosity ($L_b$) evolution as $(1+z)^\gamma$.  For the former we show the results for $\delta=0.2$ as an example.  The results for $\delta=0.5$ and $0.8$ are similar, showing inconsistency with the BATSE and many \textit{Swift} $\log N-\log P$ constraints. An `N' in the significance column indicates that test fails to beyond a $3\sigma$ level. Columns 8 and 10 indicate the significance level for the BATSE and \textit{Swift} $\log N-\log P$ tests, respectively.}
\begin{tabular}{cccccccccc}
\hline
\hline
 Model & L-z? & LF parameters & z &  L & Significance & BATSE LNLP & Sig & \textit{Swift} LNLP & Sig  \\
  & Y/N & $(\alpha_1,L_b, \alpha_2)$ & Stat, P-value & Stat, P-value &  z/L & T stat, P-value &  & T stat, P-value & \\
  \textbf{Rate evolution with $z$} & \\
  \textbf{GRB rate $\propto$ SFH*$(1+z)^\delta$} & \\
  $\delta$ \textbf{=0.2} & \\
  BL & Y & (0.05,400) & - & - & - & - & N & - & 2$\sigma$ \\
          & Y & (0.18,500) & - & - & - & - & N & - & 2$\sigma$ \\
          & Y & (0.24,800) & - & - &-  & - & N & - & 2$\sigma$ \\
          & Y & (0.29,800) & - & - & - & - & N & - & 2$\sigma$ \\
  HB & N & - & - & - & - & - & - & - & - \\
  CN & N & - & - & - & - & - & - & - & - \\
  PM & N & - & - & - & - & - & - & - & - \\
  \hline
  BL+Li & N & - & - & - & - & - & - & - & - \\
  HB+Li & N & - & - & - & - & - & - & - & - \\
  CN 0.1 cut & N & - & - & - & - & - & - & - & - \\
  \hline
   BL+LN weighted  & Y & (0.54,900,2.2) &- & -& -& -& N &- & N \\
           & Y & (0.46,700,2.2) &- &- &- &- & N & -& N \\
           & Y & (0.4,500,2.2) &- & -& -&- & N &- & N \\
           & Y & (0.24,400,2.2) &- &- & -&- & N &- & N \\
  HB+LN weighted & Y & (0.54,900,2.2) & -& -&- &- & N &- & 3$\sigma$ \\
          & Y & (0.54,900,2.2) &- &- &- & -& N &- & 3$\sigma$ \\
  CN weighted & N & - & - & - & - & - & - & - & - \\
  \hline

\textbf{LF break evolution} & \\
$L_b \propto L_b *(1+z)^\gamma$ & \\
$\gamma$ \textbf{=1.0} & \\
HB & Y & (0.15,500) & 2.77368	, 0.02321  & -0.37816, 0.44065 &  3$\sigma$/ 1$\sigma$  & 8.98113, 0.00015  & N &  2.04678, 0.04582 & 3$\sigma$ \\
CN & Y & (0.05,300) & 0.9696, 0.13296  & -0.66268, 0.5273 & 2$\sigma$/ 1$\sigma$ & 0.13296, 0.29664 & 2$\sigma$ & -0.00699, 0.33353 & 1$\sigma$ \\
& Y & (0.09,400) & 1.33265, 0.0926 & -0.49987, 0.47758 & 2$\sigma$/ 1$\sigma$ & 2.86931, 0.0213 & 3$\sigma$ & 0.64614, 0.18262 & 2$\sigma$ \\
& Y & (0.23,500) & 2.99502, 0.01906 & 0.24022, 0.27003 & 3$\sigma$/ 1$\sigma$ & -0.59966, 0.50807 & 1$\sigma$ & 0.50419, 0.20924 & 2$\sigma$ \\
& Y & (0.23,800) & 1.94944, 0.05035 & 1.81983, 0.05714 & 2$\sigma$/ 2$\sigma$  & 9.08464, 1.40E-04 & N & 1.49638, 0.07867  & 2$\sigma$ \\
& Y & (0.16,600) & 0.97282, 0.13253 & -0.03763, 0.34191 & 2$\sigma$/ 1$\sigma$ & 6.35583, 0.00124 & N & 2.12331, 0.04256 & 3$\sigma$ \\
\hline

$\gamma$ \textbf{=1.1} & \\
CN & Y & (0.13,300) & 1.1944, 0.1063 & -0.60168, 0.50868 & 2$\sigma$/ 1$\sigma$ & -0.10903, 0.36181 & 1$\sigma$ & 0.31854, 0.25158 & 2$\sigma$ \\
\hline

$\gamma$ \textbf{=1.2} & \\
HB & Y & (0.12,300) & 3.43185, 0.01311  & -0.86134, 0.58714 & 3$\sigma$/ 1$\sigma$ & 5.13402, 0.00331 & 3$\sigma$ & 0.90169, 0.14221 & 2$\sigma$ \\
CN & Y & (0.23,500) & 1.63419, 0.06861 & 1.25677, 0.09989 & 2$\sigma$/ 2$\sigma$ & 3.68355, 0.01067  & 3$\sigma$ & 0.32701, 0.24697 & 2$\sigma$ \\
& Y & (0.08,200) & 1.66239, 0.06672 & -1.02453, 0.63452 & 2$\sigma$/ 1$\sigma$  & -0.58847, 0.50465 & 1$\sigma$ & 1.26917, 0.09866 & 2$\sigma$ \\
\hline

$\gamma$ \textbf{=1.3} & \\
CN & Y & (0.12,200) & 1.33772, 0.09214 & -0.33412, 0.42743 & 2$\sigma$/ 1$\sigma$ & -0.65832, 0.52597 & 1$\sigma$ & 0.1113, 0.3022 & 2$\sigma$ \\
\hline

$\gamma$ \textbf{=1.4} & \\
CN & Y & (0.17,300) & 1.61382, 0.07001 & 0.82904, 0.15277 & 2$\sigma$/ 2$\sigma$ & 1.65549, 0.06718 & 2$\sigma$ & 1.00256, 0.12868 & 2$\sigma$ \\
\hline

$\gamma$ \textbf{=1.5} & \\
CN & Y & (0.23,200) & 1.73898, 0.06187 & 0.60585, 0.18986 & 2$\sigma$/ 2$\sigma$ & -0.82308, 0.57576  & 1$\sigma$ & 0.14452, 0.2937 & 2$\sigma$ \\
\hline

\hline
\end{tabular}
\end{table*}
\label{lastpage}

\end{document}